\documentclass[aps,twocolumn,showpacs]{revtex4}

\usepackage{graphicx}
\usepackage{amsmath}
\usepackage{hyperref}

\newcommand{\abs}[1]{\left| #1 \right|}
\newcommand{\tn}[1]{\textnormal{#1}}
\newcommand{\gcmcub}{{\rm g}~{\rm cm}^{-3}}
\newcommand{\msun}{{M_{\odot}}}

\newcommand{\rovel}{\rho_{24}v_{250}^{-1}}
\newcommand{\lsim}{\stackrel{\textstyle <}{_\sim}}

\begin{document}

\title{Impact of strange quark matter nuggets on pycnonuclear
  reaction rates in the crusts of neutron stars}

\date{\today}

\author{B.~Golf} \email{{bsouci@gmail.com}} \affiliation{Department of
  Physics, San Diego State University, 5500 Campanile Drive, San
  Diego, California 92182, USA}

\author{J.~Hellmers} \email{hellmersjl@mac.com}
\affiliation{Department of Physics, San Diego State University, 5500
  Campanile Drive, San Diego, California 92182, USA}

\author{F.~Weber} \email{fweber@sciences.sdsu.edu}
\affiliation{Department of Physics, San Diego State University, 5500
  Campanile Drive, San Diego, California 92182, USA}

\begin{abstract}{This paper presents an investigation into the
    pycnonuclear reaction rates in dense crustal matter of neutron
    stars contaminated with strange quark matter nuggets. The presence
    of such nuggets in the crustal matter of neutron stars would be a
    natural consequence if Witten's strange quark matter hypothesis is
    correct. The methodology presented in this paper is a recreation
    of a recent representation of nuclear force interactions embedded
    within pycnonuclear reaction processes.  The study then extends
    the methodology to incorporate distinctive theoretical
    characteristics of strange quark matter nuggets, like their low
    charge-per-baryon ratio, and then assesses their effects on the
    pycnonuclear reaction rates. Particular emphasis is put on the
    impact of color superconductivity on the reaction rates. Depending
    on whether or not quark nuggets are in this novel state of matter,
    their electric charge properties vary drastically which turns out
    to have a dramatic effect on the pycnonuclear reaction
    rates. Future nuclear fusion network calculations may thus have
    the potential to shed light on the existence of strange quark
    matter nuggets and on whether or not they are in a color
    superconducting state, as suggested by QCD.}
\end{abstract}


\maketitle

\section{Introduction}

First proposed in 1934 and subsequently observed in 1967, neutron
stars are the remnants of supernova explosions of aging supergiant
stars. They are extremely massive at approximately 1.4 times the mass
of the sun. The interior structure of a neutron star is still the
subject of debate, as neutron stars are tremendously compact objects,
crushing their mass into a radius of roughly ten kilometers.  This
makes them some of the densest objects known to man, with core
densities possibly more than ten times the density of atomic nuclei.
This extreme compression provides a high-pressure environment in which
numerous subatomic particle processes are believed to compete with
each other and novel phases of matter may exist. The most spectacular
ones stretch from the generation of hyperons and baryon resonances
\cite{glen97:book,weber99:book}, to quark deconfinement
\cite{ivanenko65:a,fritzsch73:a,baym76:a,keister76:a,%
  chap77:a+b,fech78:a,glen91:pt,alford07:nature}, to the formation of
boson condensates
\cite{kaplan86:a,brown87:a,brown95:a,li97:ab,brown97:a}.  (For
overviews, see \cite{glen97:book,weber99:book,lattimer01:a,weber05:a,%
  klaehn06:a,sedrakian07:a,page06:a}.)  It has also been suggested
that strange quark matter \cite{farhi84:a,schaffner97:a}, made up of
roughly equal numbers of up, down and strange quarks, may be more
stable than ordinary atomic nuclei \cite{witten,bodmer,terazawa}. This
intriguing possibility is known as the strange quark matter
hypothesis. In the latter event, neutron stars could be made almost
entirely of strange quark matter rather than confined hadronic matter
\cite{alcock86:a,alcock88:a,madsen98:b}.  If quark matter exists in
neutron stars it ought to be a color superconductor
\cite{rajagopal01:a,alford01:a,alford98:a,rapp98+99:a}. This
fascinating possibility has renewed tremendous interest in the physics
and astrophysics of quark matter
\cite{weber05:a,page06:a,rajagopal01:a,alford01:a,alford08:a}.

The idea that quark matter may exist in neutron stars is not new
\cite{ivanenko65:a,fritzsch73:a,baym76:a,keister76:a,%
  chap77:a+b,fech78:a,glen91:pt}. With the dissolution of nuclei in
the cores of neutron stars, the crush of gravity may allow even
protons and neutrons to be broken into their constituent
components. The baryons at this point exist in such close proximity
that their quarks are effectively free. With up and down quarks being
the least massive, they would be the first to appear in this
deconfined state.  Furthermore, given the relatively small mass of the
strange quark, it is possible that highly energetic up and down quarks
from nucleons may convert to low-energy strange quarks at about the
same density \cite{glen91:pt}. This mixture of up, down, and strange
quarks is referred to as strange quark matter if its energy per baryon
is less than that of ordinary nuclear matter, as expressed by the
strange quark matter hypothesis. Another distinguishing feature
between nuclear matter and strange matter is that the latter is
self-bound and electrically charge neutral. For that reason strange
matter objects can exist stably for a tremendous range of baryon
numbers, $A$, ranging from $\sim 10^2$ to $10^{57}$. The low
baryon-number end is formed by strangelets (strange nuggets)
\cite{farhi84:a,berger87:a,schaffner92:a,gilson93:a,zhang02:a} , while
the high baryon-number end is populated by strange (quark matter)
stars
\cite{alcock86:a,alcock88:a,madsen98:b,weber05:a,glen97:book,weber99:book}.

If strange quark matter exists, it may be present in any of the
neutron star regions -- in the outer and inner crust through accretion
from a companion compact stellar object and in the core through state
changes.  The base of the outer crust is characterized by a transition
known as neutron drip, where the attractive nuclear force that holds
subatomic particles together in nuclei is saturated and neutrons
subsequently leak out of the nuclei. The inner crust is a composite of
these free neutrons and a lattice of neutron-rich matter, making this
region a suitable environment for the study of high-density nuclear
interactions, known as pycnonuclear reactions when a lattice is
involved.  The lattice structure is a critical feature, as the extreme
gravitational force of neutron stars can offset the kinetic and
potential energies of the constituent matter in the crust of neutron
stars, with the ground state lattice configuration balancing the
competing forces.  Within this framework, pycnonuclear reactions are
distinctive in that they can occur at very low temperatures.  In the
inner crust, it is possible that pycnonuclear reaction rates between
SQM and normal atomic matter may be very different from those between
ordinary nuclei. Along this line of inquiry, this is a first step in
the examination of pycnonuclear reaction rates between normal nuclear
matter and a simple model incorporating some distinct characteristics
of strange quark matter.

To this end, the current undertaking recreates recent efforts to
quantify pycnonuclear reaction rates, efforts that included ion-ion
interaction models systematically constructed from nucleon-nucleon
interactions. This study then widens the scope, allowing for the
inclusion of theoretical characteristics of strange quark matter. The
final encoded methodology calculates rudimentary pycnonuclear reaction
rates for different species of inner crust matter in neutron stars,
taking basic features of SQM and normal matter and providing reaction
rates between them. No assumptions are made about the relative amount
of strange quark matter present in the inner crust, beyond postulating
its existence; the pycnonuclear reaction calculations are limited to a
single primitive lattice cell, with a single strangelet surrounded by
normal matter ions.  Future work may leverage the magnitude of the
differences in reaction rates to modify solar heating models for
neutron stars, perhaps providing clarification on the existence of
strange quark matter.

The paper is organized as follows. Section \ref{sec:pycno} provides an
introduction into nuclear reaction rates in high-density environments.
The characteristics of strange quark matter that appear to affect the
pycnonuclear reaction rates are presented in Section \ref{sec:sqm}.
This is followed by a review of observed trends and an exploration of
possible future work presented in Section \ref{sec:conclusions}.

\section{Pycnonuclear Reaction Rates}\label{sec:pycno}

Generally speaking, in pycnonuclear reaction rates, the incoming and
target particle densities usually considered in a nuclear reaction are
replaced by geometric crystal lattice calculations that account for
the separation distance between a specific number of nearby, bound
ions. The kinetic energy, usually involving a calculation of the
relative velocity, is subsumed in an analysis of the total potential
energy of a reacting pair of particles, stemming from both the
configuration of the lattice and the vibration of the bound ions. In
turn, the incident flux through a pure coulomb barrier is replaced by
an approximate calculation of the wave function transmitted through
the total, screened, effective barrier between the two ions.

In a high-density environment, such as a neutron star, it is expected
that matter will settle into a low-energy configuration, such as the
body-centered cubic which is assumed to have the greatest binding
energy per nucleus \cite{peierls}.  The repulsive framework of
similarly-charged ions enveloping two reacting ions in a bcc lattice
reduces the Coulomb barrier between them, resulting in a barrier
potential that is suppressed in both magnitude and range when compared
to that denoted by a pure Coulomb law \cite{clayton}.  Since ions
trapped in such a lattice will continue to fluctuate about their
equilibrium positions in their ground state, the ions may be able to
penetrate the Coulomb barrier of a nearby ion \cite{shapiro}.  In this
effort, consistent with the most common treatment of pycnonuclear
reactions within the model of the bcc lattice, the ``static lattice''
approximation - in which all the surrounding ions as well as the
center of mass of the reacting ion pair are considered frozen at their
equilibrium positions - is used \cite{salpeter,ocppycno}.
Furthermore, the electrons are assumed to act as a background electron
gas with a density consistent for charge neutrality, moving within the
arrangement of positive charge.

Addressing the units of the reaction rate calculation, lengths and
energies for one-component (OCP) and multi-component plasmas (MCP) are
usually measured in terms of the characteristic quantities $r^*$ and
$E^*$, respectively. These terms are defined as
\begin{equation}
r_{\tn{OCP}}^* = \frac{\hbar^2 4 \pi \epsilon_0}{MZ^2 e^2},
\end{equation}
\begin{equation}
E_{\tn{OCP}}^* = \frac{Z^2 e^2}{r_{\tn{OCP}}^*},
\end{equation}
\begin{equation}
r_{\tn{MCP}}^* = \frac{A_1+A_2}{2A_1A_2 Z_1Z_2}
\frac{\hbar^2 4 \pi \epsilon_0}{H e^2},
\end{equation}
\begin{equation}
E_{\tn{MCP}}^* = \frac{Z_1Z_2 e^2}{r_{\tn{MCP}}^*},
\end{equation}
in analogy with the Bohr radius and the ground state energy of an
electron in the Bohr model for hydrogen \cite{salpeter}. Here, {\em M}
is twice the reduced mass for a pair of nuclei, and {\em H} is the
atomic mass unit in grams. The density of the plasma can then be
expressed in terms of $r^*$ and a dimensionless inverse-length
parameter $\lambda$, introduced by Salpeter and Van Horn in their
seminal work on pycnonuclear reactions and defined as
\begin{eqnarray}
  \lambda_\tn{OCP} &\equiv& r_\tn{OCP}^* \left( \frac{N_\tn{A}}{2} 
  \right)^{1/3} \nonumber \\
  &=& \frac{1}{AZ^2} \left( \frac{1}{A} \frac{\rho}{1.3574 \times 10^{11} 
\tn{g cm}^{-3}} \right)^{1/3},
\end{eqnarray}
\begin{eqnarray}
  \lambda_\tn{MCP} &\equiv& r_\tn{MCP}^* \left( \frac{N_\tn{E}}{2Z} 
  \right)^{1/3} \nonumber \\
  &=& \frac{A_1+A_2}{2A_1A_2 Z_1Z_2 \left( Z_1+Z_2 \right)^{1/3}} \nonumber \\
  &\times&\left( \frac{\langle Z \rangle}{\langle A\rangle} 
    \frac{\rho}{1.3574 \times 10^{11} \tn{g cm}^{-3}} \right)^{1/3},
\end{eqnarray}
where $\langle$Z$\rangle$ and $\langle$A$\rangle$ are the mean charge
and mean mass number \cite{salpeter, mcppycno}.

With respect to the density model that describes the nuclear densities, a 
two-parameter Fermi (2pF) distribution,
or Woods-Saxon potential, is used:
\begin{equation}
  \rho \left(r\right) = \frac{\rho_0}{1 + \exp\left( \frac{r-R_0}{a} \right)}.
\end{equation}
The normalization condition is
\begin{equation}
4\pi = \int_0^\infty \rho\left( r \right) r^2 dr = X,
\end{equation}
where $X$ could be the number of protons, neutrons, or nucleons used
to determine $\rho_0$ \cite{globalbp}.  This distribution can recreate
the saturation of the nuclear medium and the rapid fall-off that
brings out the notion of the radius $R_0$ of the nucleus
\cite{globalbp}.

On the whole, as an input to the 2pF distributions, the radii of most
nuclei are well described by
\begin{equation}
\tn{radius (in fermi)} = 1.31 \times A^{1/3} - 0.84,
\end{equation}
where {\em A} is the number of nucleons in a given nucleus
\cite{bpdatachk}.  The rapid decrease in density is tied to the
diffuseness parameter {\em a}.  This parameter plays a key role in
subsequent models that vary the range of the nuclear force.  Both
nucleon and matter densities, coupled with their average diffuseness
parameters $\bar{a}_\tn{nucleon}$ = 0.50 fm and $\bar{a}_\tn{matter}$
= 0.56 fm, respectively, give similar radii \cite{braz}.

\subsection{Long-Range Interactions}

The constituent calculations of a reaction rate are broken up by where
specific forces are dominant, with this factoring into long-range and
short-range interactions possible because the lattice separations are
much greater than the radii of the ions.  As mentioned earlier,
calculations for the total kinetic energy involved in normal
thermonuclear reactions are replaced by a calculation of the energies
of the bound ions in pycnonuclear reactions. Beginning with the
electrostatic lattice energy that maintains the structure of the ions,
an exact expression for the electrostatic screening potential is
available, involving a six-fold sum of the locations of all
nearest-neighbor nuclei in the lattice \cite{carr}. However, an
approximation based on the static lattice paradigm is also possible,
reducing the six-fold sum to a triple lattice sum based on the
separation distance $r$ between the reacting ion pair. This
approximation leads to the total electrostatic interaction energy per
nucleus being given by
\begin{equation}
E_\tn{Coulomb}^\tn{bcc} = 1.81962 \frac{Z^2 e^2}{a} =
1.81962\ \lambda E^*,
\end{equation}
where {\em a} is a constant of the lattice derived from the baryon
number density \cite{carr}.  However, the lattice energy is not the
only energy to consider; the nuclei are also vibrating in their ground
states within the bcc lattice. The oscillation frequency of the nuclei
around their equilibrium lattice sites is of the order of magnitude of
the ion-plasma frequency $\omega_\tn{p}$, defined by \cite{salpeter}
\begin{equation}
  \hbar \omega_\tn{p} = \hbar \left( 4 \pi \frac{2Z^2 e^2}{a^3 AH} 
  \right)^{1/2} 
  = 2.7426\ \lambda^{1/2} E_\tn{Coulomb},
\end{equation}
with an anisotropic harmonic oscillator potential describing the
relative motion of the two ions.  The energy-level spacings for
vibrations in the $z$-axis separating two nearest-neighbor ions in the
bcc lattice are \cite{salpeter}
\begin{equation}
\hbar \omega_\tn{z} = 0.6752\ \hbar \omega_\tn{p}.
\end{equation}
This vibrational energy in the $z$-direction along the axis connecting
the two reacting nuclei can then be added to the electrostatic energy
per particle in the bcc lattice to determine the total energy of the
pair-wise nuclear reaction.

To determine the wavefunction of the incoming particle using the
total, screened Coulomb potential, Salpeter and Van Horn used a
Wentzel-Kramers-Brillouin (WKB) approximation to show that
\begin{eqnarray}
  \psi \left(r\right) &=& 0.553\ \frac{\lambda^{7/8}}{\left( r^*\right)^{3/2}} 
  \left( r_\tn{n}\right)^{-1/4}  \nonumber \\
  &\times& \exp \left[ -\frac{1}{2} \lambda^{-1/2} 
    \left( J - \lambda^{1/2} K \right) + 2r_\tn{n}^{1/2} \right],
\end{eqnarray}
with
\begin{equation}
  J - \lambda^{1/2} K = 2.638 - 3.6\ \lambda^{1/2},
\end{equation}
is the wave function evaluated at the nuclear radius for the relative
motion of the reacting nuclei in the static approximation of the
lattice \cite{salpeter}.  At this point, all of the long-range
interactions for the pair-wise reaction rate $W$ are included,
resulting in
\begin{equation}
  W = \frac{8\ S\left(E \right)}{\hbar \left(r^* \right)^2} \left(r^* 
  \right)^3 r_\tn{n}^{1/2} \exp \left(4r_\tn{n}^{-1/2} \right) 
  \abs{\psi_\tn{E0} \left( r_\tn{n}\right)}^2.
\end{equation}

From here, only the eight nearest-neighbors surrounding any given
lattice point are taken into account, resulting in the following
expression for the reaction rate per unit volume:
\begin{equation}
  P = \frac{8\ \rho}{2\ \langle A \rangle H} \ W\  \tn{reactions cm}^{-3} ~ 
  \tn{s}^{-1}.
\end{equation}
Possible interactions with all other nuclei are ignored.  The
temperature-independent pycnonuclear reaction rate for a bcc lattice,
encompassing all the long-range effects of a pycnonuclear reaction,
is thus given by
\begin{eqnarray}
  P &=& 8\ \frac{\rho A_1 A_2 Z_1^2 Z_2^2}{A_1 + A_2} S\left(E \right) 
  \lambda^{7/4} \nonumber \\
  &\times& \exp \left( \frac{-2.638}{\lambda^{1/2}} \right)
  \left( 3.90 \times 10^{46} ~ \tn{s}^{-1} \right),
\end{eqnarray}
where $\rho$ is in g$\cdot$cm$^{-3}$ and the S-factor is in units of
MeV$\cdot$barns \cite{salpeter,mcppycno}.

\subsection{Astrophysical S-Factor}

The calculations up to this point have started from a macroscopic
visualization of the long-range effects.  The calculations that follow
will invert this viewpoint, instead constructing the astrophysical
S-factor from separate models of the different forces with short-range
effects.  With respect to units in S-factor calculations, the $r^*$
length scale of the bcc lattice is dropped for the more convenient use
of fermi.  One begins with the total effective potential, the sum of
the nuclear and unscreened Coulomb components \cite{ocppycno}:
\begin{equation}
  V_\tn{eff}\left( r,E \right) = V_\tn{N} \left( r,E \right) + V_\tn{C}
  \left( r \right) + \frac{ l \left( l+1 \right) \hbar^2 } 
  {2 \mu r^2}.
\end{equation}
Since the dominant interaction at the low energies of pycnonuclear
reactions is via spherically-symmetric {\em s}-wave scattering, {\em l} is
set to zero in all calculations presented in this effort
\cite{salpeter}.

The first term, the nuclear potential, is based on one of several
methods created in a recent model, referred to in this work as the Sao
Paulo model, which was developed by a joint collaboration from several
universities and institutes in Brazil
\cite{globalbp,paulinonloc,lett79, c58}.  One starts with the concept
that the effective one-body interaction between two nuclei can be
written schematically as
\begin{eqnarray}
  V\left(\vec{r},\vec{r}\,' \right) &=& V_\tn{bare} \left(\vec{r},\vec{r}\,' 
  \right) \nonumber \\
  &+& \sum_i V_i \left(\vec{r}\right) G_i \left(\vec{r},\vec{r}\,'; E \right) 
  V_i\left(\vec{r}\, ' \right)  .
\end{eqnarray}
The first term on the right is the bare interaction, representing the
ground state of an interaction between two nuclei.  The second
highly-energetic term encompasses polarization and intermediate
states, like inelastic channels \cite{paulinonloc}.  This second term
is not considered in this effort.

The first term is the focal point of the Sao Paulo model and can be
calculated using multiple methods, though the one presented here
associates an energy dependence of the derived nuclear potential with
nonlocal quantum effects.  It begins by concentrating on the effects
arising from the Fermi nature of the nucleons.  When calculating
interaction potentials between nuclei, these effects translate into a
nonlocality.  The nonlocality of the bare interaction is solely due to
the Pauli exclusion principle, hereafter referred to as Pauli
nonlocality \cite{paulinonloc}.  This Sao Paulo methodology creates a
local equivalent potential that makes use of a double-folding
methodology whereby a representation of the effective nucleon-nucleon
force is folded, i.e., integrated, with the densities of the
interacting nuclei:
\begin{equation}
  V_\tn{bare} = V_\tn{folding} = \int \rho_1 \left(r_1\right) \rho_2 \left( 
    r_2 \right) \upsilon_\tn{NN} \ d\vec{r}_1 \ d\vec{r}_2 \, .
\end{equation}
The result is dependent only on the number of nucleons in the nuclei.

The Sao Paulo group adopts the following form for the bare nuclear
potential between two nuclei:
\begin{equation}
  V_\tn{bare} \left( \vec{r}, \vec{r}\, ' \right) = V_\tn{NL} \left( \frac{ 
      \abs{\vec{r} + \vec{r}\, '}}{2} \right)
  \exp \left[-\frac{ \left( \vec{r} - \vec{r}\, ' \right)^2}{b^2} \right],
\label{eq:24}
\end{equation}
where {\em b} is the range of the Pauli nonlocality of the ion-ion
interaction \cite{paulinonloc}.  In a sense, this term measures the
range of the nonlocal effect where the nuclear force is operative.
Jackson and Johnson explain in previous work that
\begin{equation}
b \approx \frac{b_0 m}{\mu}
\end{equation}
within their single folding model, where $b_0$ is the nucleon-nucleus
nonlocality parameter, {\em m} is the nucleon mass, and $\mu$ is the
reduced mass of the nucleus-nucleus system \cite{bparam}.  Fitting to
nucleon-nucleus scattering data, Perey and Buck later show that $b_0 =
0.85\ \tn{fm}$ \cite{perey}.  Continuing to use the density-folding
formalism, the nonlocal potential $V_\tn{NL}$ in Eq.\ (\ref{eq:24}) is
written as \cite{lett79}
\begin{equation}
  V_\tn{NL} = V_\tn{folding} = \int \rho_1 \left(r_1\right) \rho_2 \left( r_2 
  \right) \upsilon_\tn{NN} \ d\vec{r}_1 \ d\vec{r}_2\, .
\end{equation}

Using the optical model, the bare nuclear potential is rewritten in
terms of a central potential.  This requires extracting a
local-equivalent potential from a nonlocal expression.  A local
equivalent potential is defined as \cite{perey}
\begin{equation}
  \int V \left( \vec{r},\vec{r}\, ' \right) \psi_\tn{E} \left( \vec{r}\, ' 
  \right)  d\vec{r}\, ' \equiv V \left( \vec{r}, E \right) \psi_\tn{E} 
  \left( \vec{r}  \right).
\end{equation}
Accordingly, in the central potential notation of the optical model,
the total nuclear interaction can be rewritten as \cite{braz}
\begin{eqnarray}
  \frac{1}{u_l \left( R \right)} \int_0^\infty V_l \left(R, R' \right) u_l 
  \left(R' \right) &=& V_\tn{LE} \left( R,E \right) \nonumber \\
  &+& i \, W_\tn{LE} \left(R,E\right) \, .
\end{eqnarray}
As can be seen, there is an energy-dependent, local equivalent
potential in the above equation for every term in the effective
one-body interaction, with the first term being predominantly real and
equivalent to the nonlocal, bare, nuclear interaction while the second
term is complex (and still ignored).  In this model, the energy
dependence of $V_\tn{LE} \left(R,E \right)$ is chiefly due to Pauli
nonlocality in the bare nuclear interaction $V_\tn{bare}
\left(\vec{r}, \vec{r}\,' \right)$.

To create this local equivalent potential, the Sao Paulo group uses
the Perey and Buck prescription, namely that for a given non-local
optical potential depth $(V_\tn{NL})$, the corresponding local
potential depth $(V_\tn{LE})$ at each energy should be chosen to
satisfy the approximate relation
\begin{equation}
  V_\tn{LE} \ \exp \left[ \frac{Mb^2}{2\hbar^2} \left(E-V_\tn{LE} \right) 
  \right] = V_\tn{NL} \, ,
\end{equation}
where {\em M} is the mass \cite{perey}. Using Eq.\ (\ref{eq:24}) and
recognizing that $V_\tn{NL}$ determines the nonlocal potential depth
for the bare nuclear interaction, the local equivalent expression in
this case can be written as the following \cite{paulinonloc}:
\begin{eqnarray}
  V_\tn{LE} \left(R,E \right) &\approx& V_\tn{F} \left(R\right) {\rm e}^{-\gamma
    \left[ E -V_\tn{C} \left(R\right) - V_\tn{LE} \left(R,E\right) \right]} \, ,
  \\
  \gamma &=& \frac{\mu b^2}{2\hbar^2}.
\end{eqnarray}

This expression has multiple possible interpretations.  In one interpretation, 
the exponent can be tied classically to the kinetic energy {\em E} 
and relative speed $\upsilon$ between the nuclei by
\begin{equation}
  \upsilon^2 = \frac{2}{\mu} E_\tn{K} \left(R\right) = \frac{2}{\mu} 
  \left[ E -V_\tn{C} \left(R\right) -V_\tn{LE} \left(R,E\right) \right] \ ,
\end{equation}
resulting in
\begin{eqnarray}
  V_\tn{LE} \left(R,E\right) & \approx & V_{\tn{F}} \left(R \right) 
  {\rm e}^{-\left[ m_{\tn{0}} b_{\tn{0}} v / 2\hbar \right]^{2}} \nonumber \\
& \approx & V_{\tn{F}} {\rm e}^{-4v^2/c^2},
\end{eqnarray}
where {\em c} is the speed of light.  This interpretation equivalently
relates the effect of Pauli nonlocality to that of a
velocity-dependent nuclear interaction.  In the current work, the
local-equivalent potential is instead related directly to the folding
potential, which incorporates an effective nucleon-nucleon interaction
dependent on the relative speed $v$ between the nucleons \cite{braz},
\begin{equation}
V_{\tn{LE}} \left( R, E\right) = V_{\tn{F}}
\end{equation}
\begin{equation}
  V_{\tn{F}} = \int \rho_{\tn{1}} \left(r_{\tn{1}} \right) \rho_{\tn{2}} 
  \left(r_{\tn{2}} \right) v_{\tn{NN}} \left( v, \vec{R} - \vec{r_{\tn{1}}} 
    + \vec{r_{\tn{2}}} \right) d\vec{r_{\tn{1}}} \ d\vec{r_{\tn{2}}}
\end{equation}
\begin{equation}
  v_{\tn{NN}} \left( v, \vec{r} \right) = v_{\tn{f}} \left( \vec{r} 
  \right) {\rm e}^{-4v^2/c^2}.
\label{eq:36}
\end{equation}
This additional modularity allows for different nucleon-nucleon
interaction models to be incorporated in future work.

\begin{figure}[htb]
\centering
\includegraphics[width=0.45\textwidth]{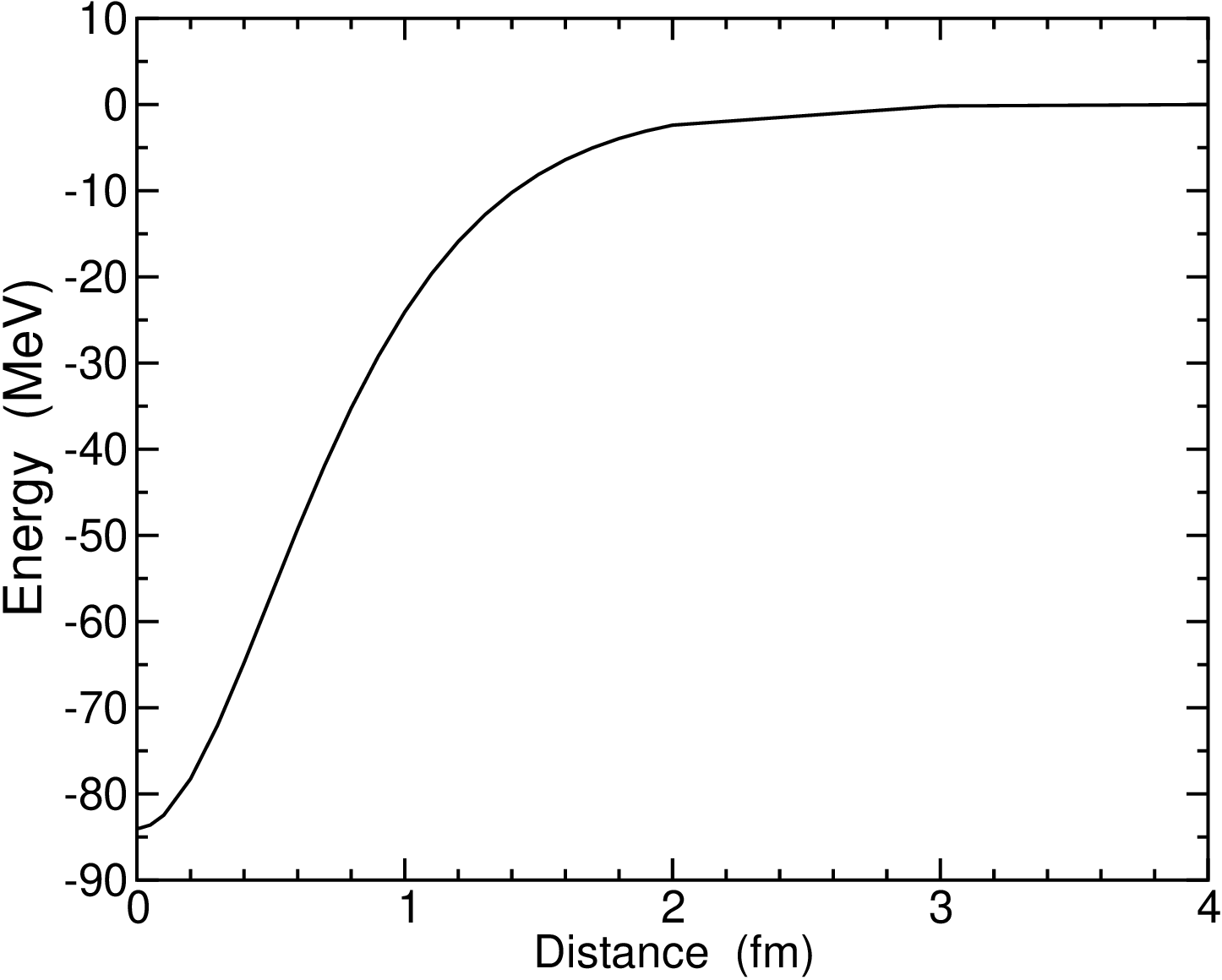}
\caption{Nucleon potential without Pauli nonlocality as
  a function of distance from nuclear center.}
\label{fig2}
\end{figure}

Accordingly, $v_{\tn{f}}\left(\vec{r}\right)$ is the next term to be
considered.  The Sao Paulo group extrapolated that the effective
nucleon-nucleon interaction could be derived by reusing a folding
potential from previous heavy-ion potential analyses,
\begin{equation}
  v_{\tn{f}} = \int \rho_{\tn{m}} \left(r_{\tn{1}} \right) \rho_{\tn{m}} \
  V_{\tn{0}} \ \delta \left( v, \vec{R} - \vec{r_{\tn{1}}} + \vec{r_{\tn{2}}} 
  \right) d\vec{r_{\tn{1}}} \ d\vec{r_{\tn{2}}},
\end{equation}
where $V_{\tn{0}} = -456\ \tn{MeV fm}^3$.  In this scheme,
$\rho_{\tn{m}}$ is the matter density of the nucleon, for which they
assumed an exponential shape,
\begin{equation}
\rho_{\tn{m}} \left(r\right) = \rho_{\tn{0}}\  {\rm e}^{-r/a_{\tn{m}}},
\end{equation}
based on electron scattering experiments that determined the intrinsic
charge distribution of the proton in free space.  Again,
$\rho_{\tn{0}}$ can be determined by the normalization condition from
the density calculations \cite{globalbp}.  The integration of
$v_{\tn{f}}$ results in a finite-range, nucleon-nucleon interaction,
\begin{equation}
  v_{\tn{f}} = \frac{V_{\tn{0}}}{64\pi a^{3}_{\tn{m}}} \ {\rm e}^{-r/a_{\tn{m}}}
  \left( 1 + \frac{r}{a_{\tn{m}}} + \frac{r^2}{3a^{2}_{\tn{m}}} \right),
\end{equation}
shown in Fig. \ref{fig2}, where the best-fit value for the matter
diffuseness $a_{\tn{m}}$ of a nucleon is 0.3~fm.

When multiplied by the velocity-dependent exponential in Eq.\
(\ref{eq:36}), effects due to Pauli nonlocality are incorporated into
this local equivalent, nucleon-nucleon interaction. Unfortunately,
with its focus on recreating nonlocal effects, this methodology does
not recreate a repulsive core for the nucleon potential, an empirical
feature that gives rise to the saturation of the nuclear
force. However, this method provides a good fit of heavy-ion potential
strengths and is similar to the M3Y potential in the surface region
\cite{globalbp}.

\begin{figure}[htb]
\centering\includegraphics[width=0.45\textwidth]{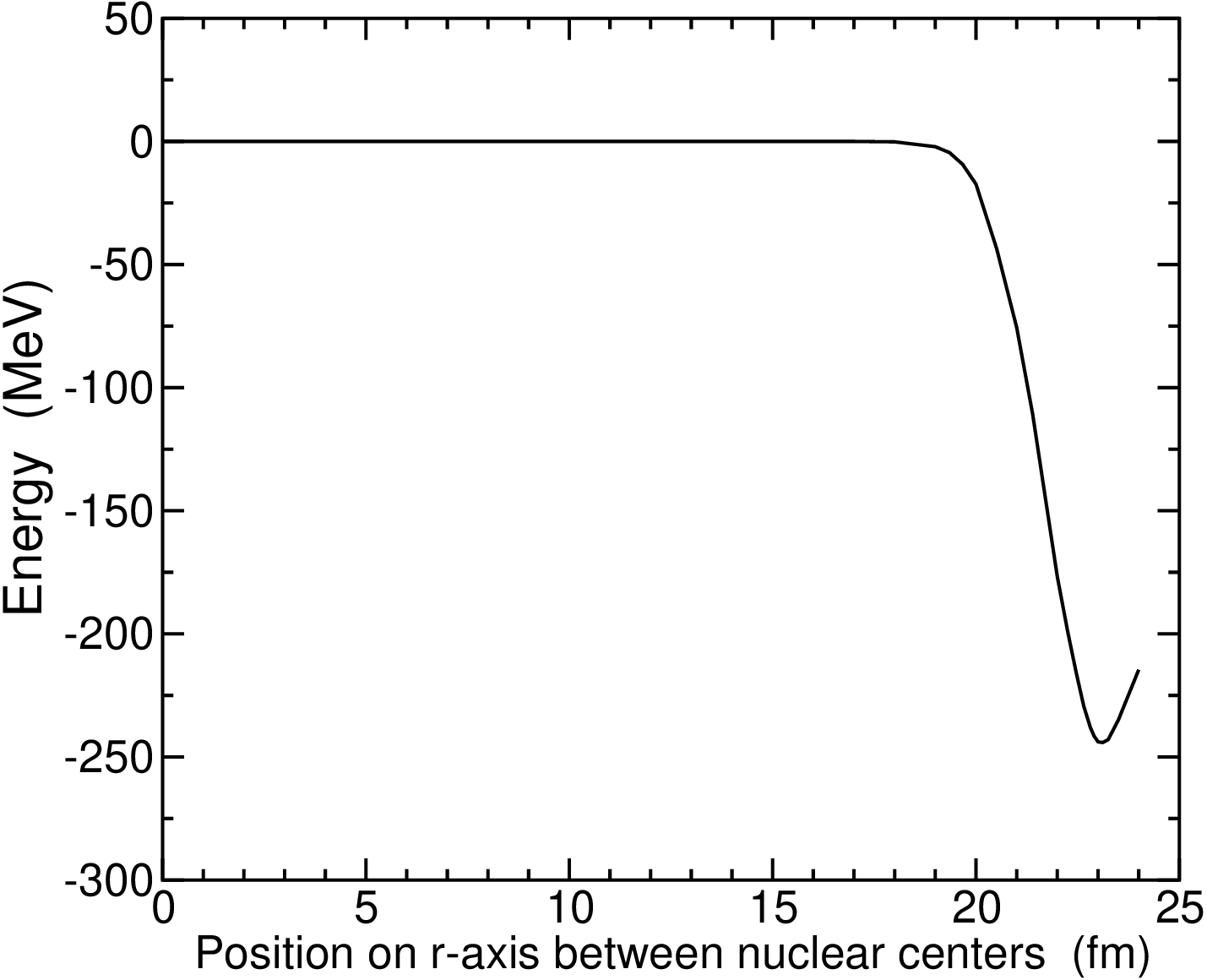}
\caption{Nuclear potential between two $\tn{C}_{12}$
  ions as a function of position on the $r$-axis between nuclear
  centers for $\rho = 10^{12}~\gcmcub$.}
\label{fig3}
\end{figure}

The nuclear potential of a system of two $\tn{C}_{12}$ ions can be
seen in Fig. \ref{fig3}. The target ion is considered locked in place
near the far-right edge of the graph. The distance between target and
projectile centers at the right edge of the graph is greater than the
actual distance between the lattice points, i.e., in a two-body
picture, the projectile has actually moved past the target; this is
the cause of the slight uptick in the nuclear potential.

Moving on from the complications of the nuclear potential, the
short-range effects of the Coulomb interaction in the effective
potential are taken into account. Outside of the nuclei, the
point-particle assumption holds.  However, in the interior of the nucleus, the expression must be
modified to account for the finite size of the nucleus, and the
potential energy due to the Coulomb repulsion becomes
\cite{paulinonloc}
\begin{equation}
  V_C\left(r\right) = \frac{\left(3R_0^2-r^2\right)Z_1 Z_2 e^2}{8\pi 
    \epsilon_0 R_0^3} \, , \qquad r<R_0,
\end{equation}
within the nucleus.
 
Having incorporated nuclear and Coulomb interactions into a total
effective potential, the calculation of the S-factor is accomplished
through the use of a barrier penetration model.  This heuristic
treatment begins with a calculation of the transmission coefficient
through the total effective potential, incorporating this into a model
based on fits of thermonuclear data to compute the nuclear cross
section \cite{ocppycno}. The definition of the S-factor,
\begin{equation}
S\left(E\right) \equiv \sigma \left(E\right) \times E \times \frac{1}{T},
\end{equation}
with the Gamow coefficient {\em T} defined for transmission through an
unscreened Coulomb barrier, can then be used to calculate the reaction
rate.

The transmission coefficient is derived by way of a WKB approximation
that requires as input the total effective potential and the total
energy.  Using the momentum of the reacting system, possible
trajectories through the total potential are orthogonal to the
surfaces of constant phase of the wavefunction \cite{schiff}. To solve
for the phase, the momentum is integrated between the classical
turning points of the effective potential, using
\begin{equation}
  \kappa (x) = \frac{1}{\hbar} \left\{ 2\mu \left[ V_\tn{eff}
  (x) - E_0 \right] \right\}^{1/2}
\end{equation}
as the momentum in one dimension \cite{schiff}. The three-dimensional
calculation for phase is then
\begin{equation}
  W\!K\!B_l = \pm \int_{r_1}^{r_2} \sqrt{\frac{8\mu}{\hbar^2} \left[ V_\tn{eff}
    \left(x\right)-E_0 \right] } \ dr.
\end{equation}
From this, the probability corresponding to the passage of the
particle through the effective potential is given by the transmission
coefficient
\begin{equation}
  T_l=\frac{\abs{\psi_\tn{trans}}^2}{\abs{\psi_\tn{inc}}^2} \approx 
  \left[ 1 + {\rm e}^{W\!K\!B_l} \right]^{-1}.
\end{equation}
In accordance with the barrier penetration model, the fusion cross
section is tied to the summation of the total particle flux
transmitted through the barrier up to the greatest value of angular
momentum in the effective potential \cite{ocppycno},
\begin{equation}
  \sigma \left(E\right) = \frac{ \pi \hbar^2}{2 \mu E_0} \ 
  \sum_{l=0}^{l_\tn{cr}}  \left(2l+1\right) T_l .
\end{equation}
The astrophysical S-factor is then calculated by multiplying $\sigma
\left(E\right)$, the total energy, and the inverse of the transmission
coefficient through a pure, unscreened Coulomb barrier.

The fact that the inverse of the Gamow penetration factor is being
used is worthy of a second look. The calculations for the total cross
section of the reaction have explicitly included a pure Coulomb
interaction in the total effective potential. It may seem as if these
efforts are being undone by removing the Gamow penetration factor from
the S-factor, but this is not the case. The inclusion of the Coulomb
interaction in the effective potential was necessary to find the
correct turning points for the WKB approximation needed to calculate
the total cross section, particularly the turning point for the onset
of the nuclear force. But, it is important to remember that the
S-factor is supposed to be an expression that compartmentalizes
nuclear force effects. Divesting the S-factor of Coulomb forces
outside of the nuclear radius follows this paradigm. Furthermore, the
reaction rate calculation separately accounts for the effects of the
more accurate, {\em screened} Coulomb potential of the bcc lattice. It
is necessary to divide out the traditional, unscreened Gamow
penetration factor in order to avoid double-counting Coulomb effects.

\subsection{Atomic Matter Pycnonuclear Reaction Rates}

\begin{figure}[htb]
\centering\includegraphics[width=0.45\textwidth]{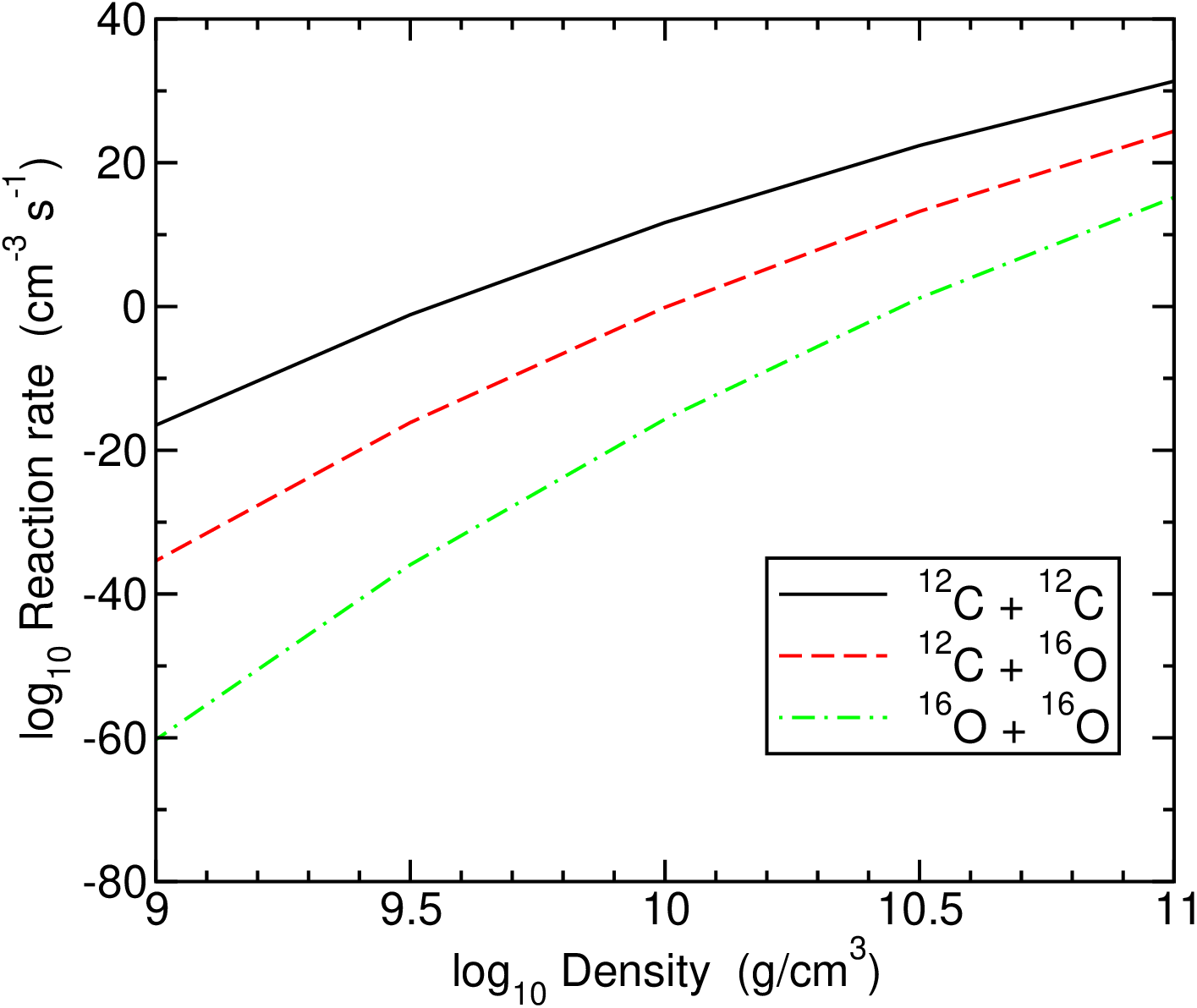}
\caption{(Color online) Reaction rates for OCP and MCP atomic matter
  as a function of density.}
\label{fig5}
\end{figure}

\begin{figure}[htb]
\centering\includegraphics[width=0.45\textwidth]{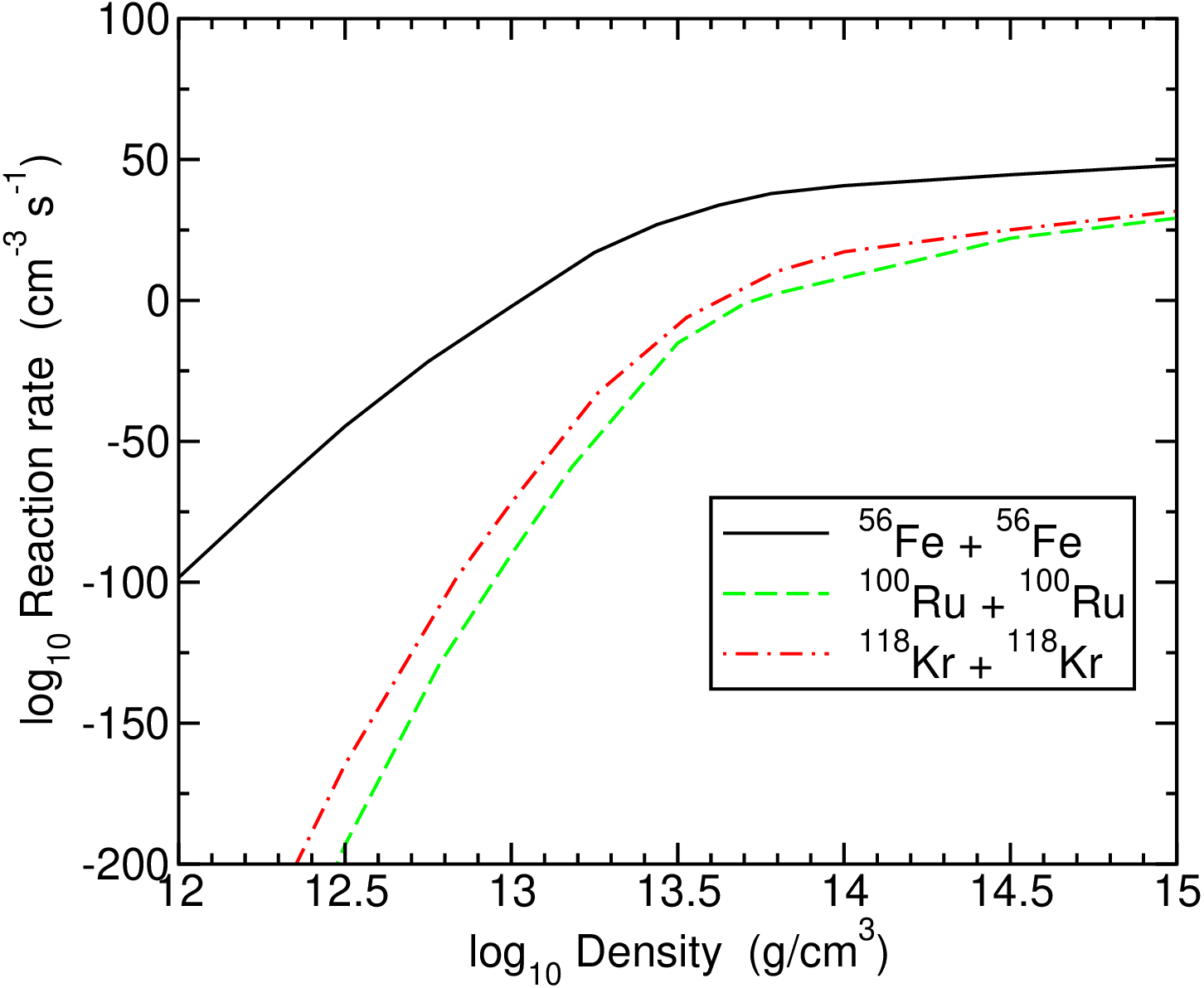}
\caption{(Color online) Reaction rates for OCP neutron-rich nuclei as
  a function of density.}
\label{fig6}
\end{figure}

With all of the constituent parts for the calculation in place,
pycnonuclear reaction rates for various elements are provided in
Figs. \ref{fig5} and \ref{fig6} for a range of densities.  The
reaction rate results for OCP and MCP nuclear matter are a good match
to those calculated by the originators of the model, an extended
collaboration that included the Sao Paulo group that created the
nuclear interaction methodology and the Joint Institute of Nuclear
Astrophysics (JINA) at the University of Notre Dame \cite{mcppycno}.

At this interim point, the pycnonuclear reaction rates confirm an
expected trend.  With the data, one can begin to see that the reaction
rates decrease as the atomic numbers of the nuclei involved
increase. When this methodology is extended to heavy ions that may be
found in the inner crust of a neutron star -- potentially iron all the
way up to neutron drip -- the reaction rates continue to plummet, as
can be seen by the rapid change in scale for Fig. \ref{fig6}. This
makes sense: The increase in the Coulomb repulsion between larger
nuclei would make it more difficult to tunnel through the barrier
between the ions, even with the effects of Coulomb screening.

\section{Strange Quark Matter}\label{sec:sqm}

From here, the effort expands to an attempt at the calculation of
pycnonuclear reaction rates between normal nuclear matter and a simple
approximation of strange quark matter. It is important to acknowledge
immediately that this extension to strange quark matter makes no
attempt to modify the methodology of the strong interaction as it has
been presented for atomic matter. However, several other models
implemented for normal nuclear matter are modified to accommodate
theoretical characteristics of strange quark matter, particularly the
bulk charge of a strangelet as well as its radius, density, and baryon
number. For the purposes of this effort, the term strangelet is used
to refer to an approximately nucleus-sized strange quark matter nugget
that replaces one of the two normal matter nuclei in the previous
reaction calculations.

\subsection{Strangelet Mass}

To replace an atomic matter nucleus within the bcc lattice with a
strangelet, one needs to know the approximate mass of the strange
quark matter nugget. There are numerical studies that have modeled the
late stages of in-spiral in binary systems of neutron stars and
massive objects such as black holes, white dwarf stars, or other
neutron stars. As tidal forces and, eventually, the actual collision
overcome the surface tension of strange quark matter and rip the
compact objects apart, the release of macroscopic lumps of matter
between $10^{-4}$ and $10^{-1}$ solar masses, or $A \propto 10^{38}$,
seems to be a characteristic feature \cite{madseng28}.  It seems a
reasonable assumption that a significant fraction of this
tidally-released material remains trapped in orbit with typical speeds
around $0.1 c$ and subject to frequent collisions. If one calculates
that the kinetic energy from these collisions is converted to
supplying the extra surface and curvature energy necessary for forming
smaller strangelets, then the released mass may end up in strangelets
with $A \approx 10^2 - 10^3$ \cite{madseng28}. This range is the focus
of the input parameters for strange quark matter baryon numbers in the
pycnonuclear reaction rate calculations presented below. Since
strangelets are self-bound, their mass-radius relationship is given by
$ M = \int_0^R \epsilon \ dV \approx 4 \pi \epsilon R^3 / 3$
\cite{madsen98:b}.  Also, instead of using the more rigorous
two-parameter Fermi model, the density is assumed to be approximately
constant throughout the strangelet.

\subsection{Strangelet Charge}

The overall charge of the strangelet is the final parameter changed,
and the one that most affects the outcome of the pycnonuclear reaction
rates in this study -- again, acknowledging that the nuclear
interaction is unchanged. We assume that the strangelets are made up
of either ordinary strange quark matter or color superconducting
strange quark matter whose condensation pattern is the
color-flavor-locked (CFL) phase. Once crucial difference between
non-CFL (NCFL) and CFL quark matter is the equality of all quark Fermi
momenta in CFL quark matter which leads to charge neutrality in bulk
without any need for electrons \cite{rajagopal01:PRL}. This has most
important consequences for the charge-to-mass ratios of strangelets.
For ordinary strangelets, the charge is approximately
\begin{equation}
Z \approx 0.1 \ m_{150}^2 \ A\, , \qquad A\ll 10^3,
\end{equation}
\begin{equation}
Z \approx 8 \ m_{150}^2 \ A^{1/3}\, , \qquad A\gg 10^3,
\end{equation}
where $m_{150} \equiv m_s / 150\ {\rm MeV}$ and $m_s$ is the mass of
the strange quark. For small {\em A}, the charge is the volume quark
charge density multiplied by the strangelet volume with a result that
is proportional to {\em A} itself. This relation holds until the
system grows larger than around 5~fm, or $A \approx 150$, at which
point the charge is mainly distributed near the strangelet surface,
and $Z \propto A^{1/3}$ \cite{madsenlett87}.  In contrast to this, the
charge-to-mass ratio of CFL strangelets is described by
\cite{madsenlett87}
\begin{equation}
  Z \approx 0.3\ m_{150} \ A^{2/3} \, .
\end{equation} 
Recently it has been shown that CFL strangelets may even have zero
electric charge if the quark pairing is strong \cite{oertel08:a}. If
strangelets should exist in the crusts of neutron stars, this
difference may provide a test of color superconductivity.

\subsection{Reaction Rates with Strange Quark Matter}

\begin{figure}[htb]
  \centering\includegraphics[width=0.45\textwidth]{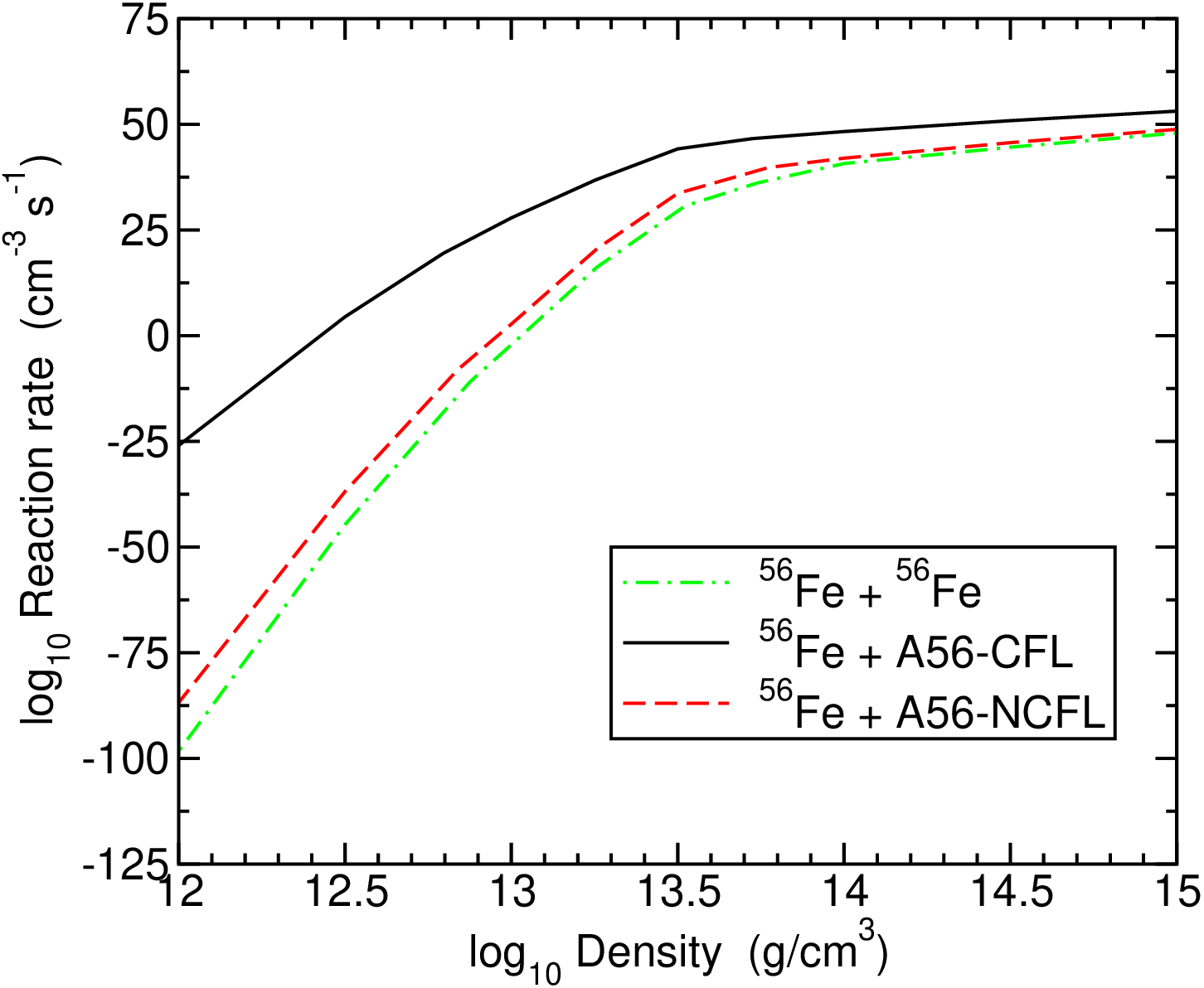}
  \caption{(Color online) Reaction rates for $^{56}$Fe and a
    strange quark matter nugget with 56 baryons as a function of
    density.}
\label{fig7}
\end{figure}

\begin{figure}[htb]
\centering\includegraphics[width=0.45\textwidth]{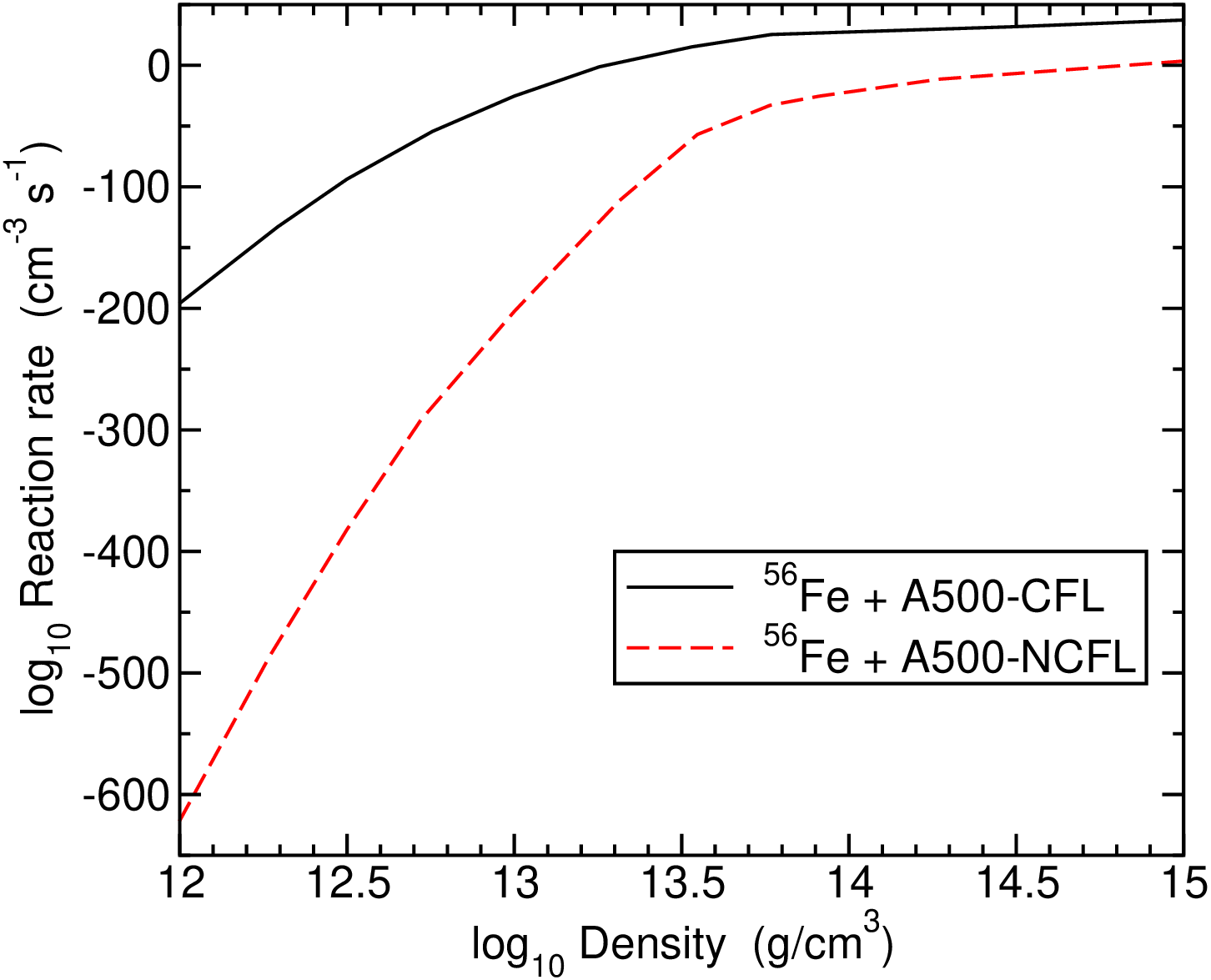}
\caption{(Color online) Reaction rates for $^{56}$Fe and a
  strange quark matter nugget with 500 baryons as a function of
  density.}
\label{fig8}
\end{figure}

\begin{figure}[htb]
  \centering\includegraphics[width=0.45\textwidth]{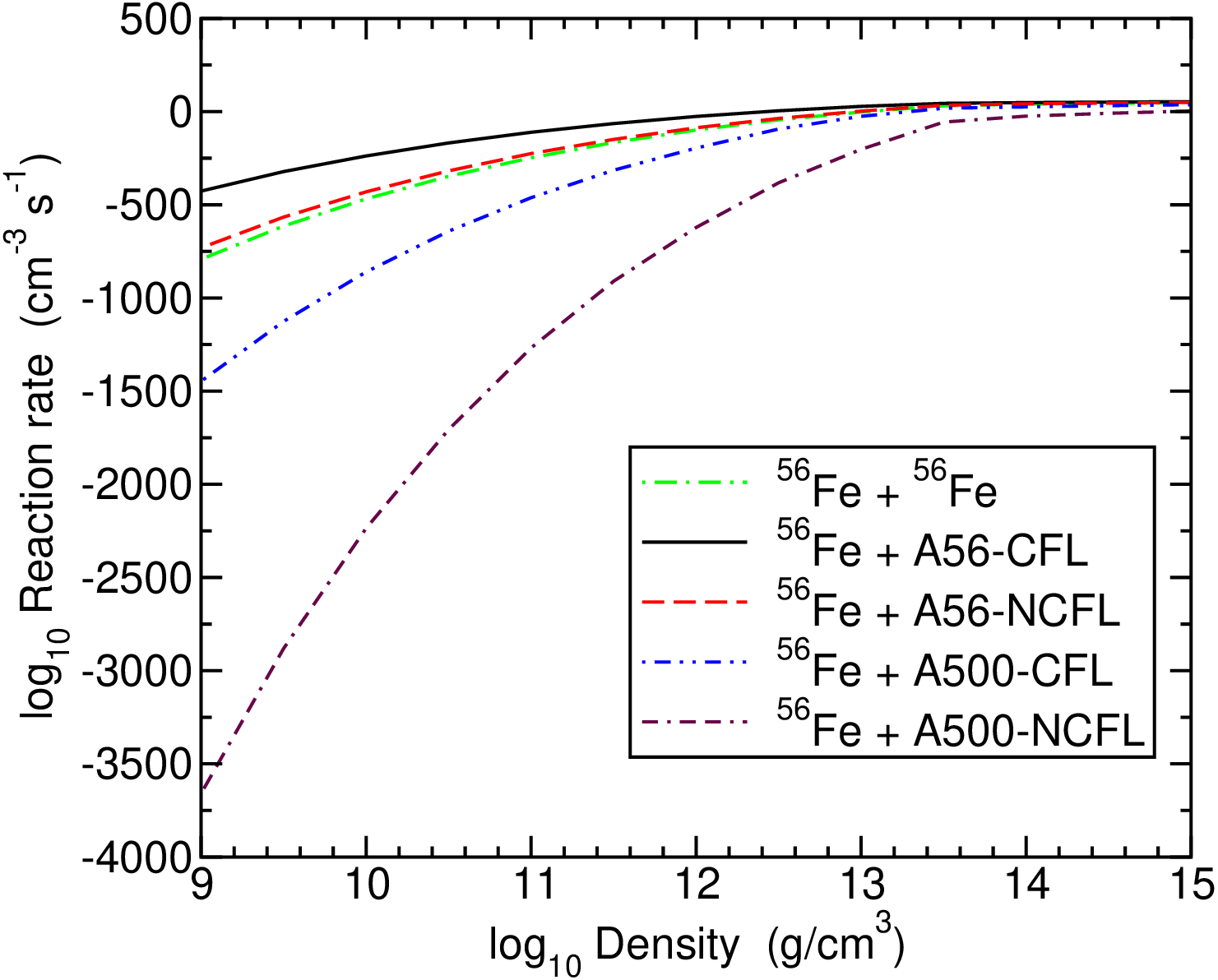}
  \caption{(Color online) All reaction rates calculated for $^{56}$Fe
    as a function of density.}
\label{fig9}
\end{figure}

\begin{figure}[htb]
  \centering\includegraphics[width=0.45\textwidth]{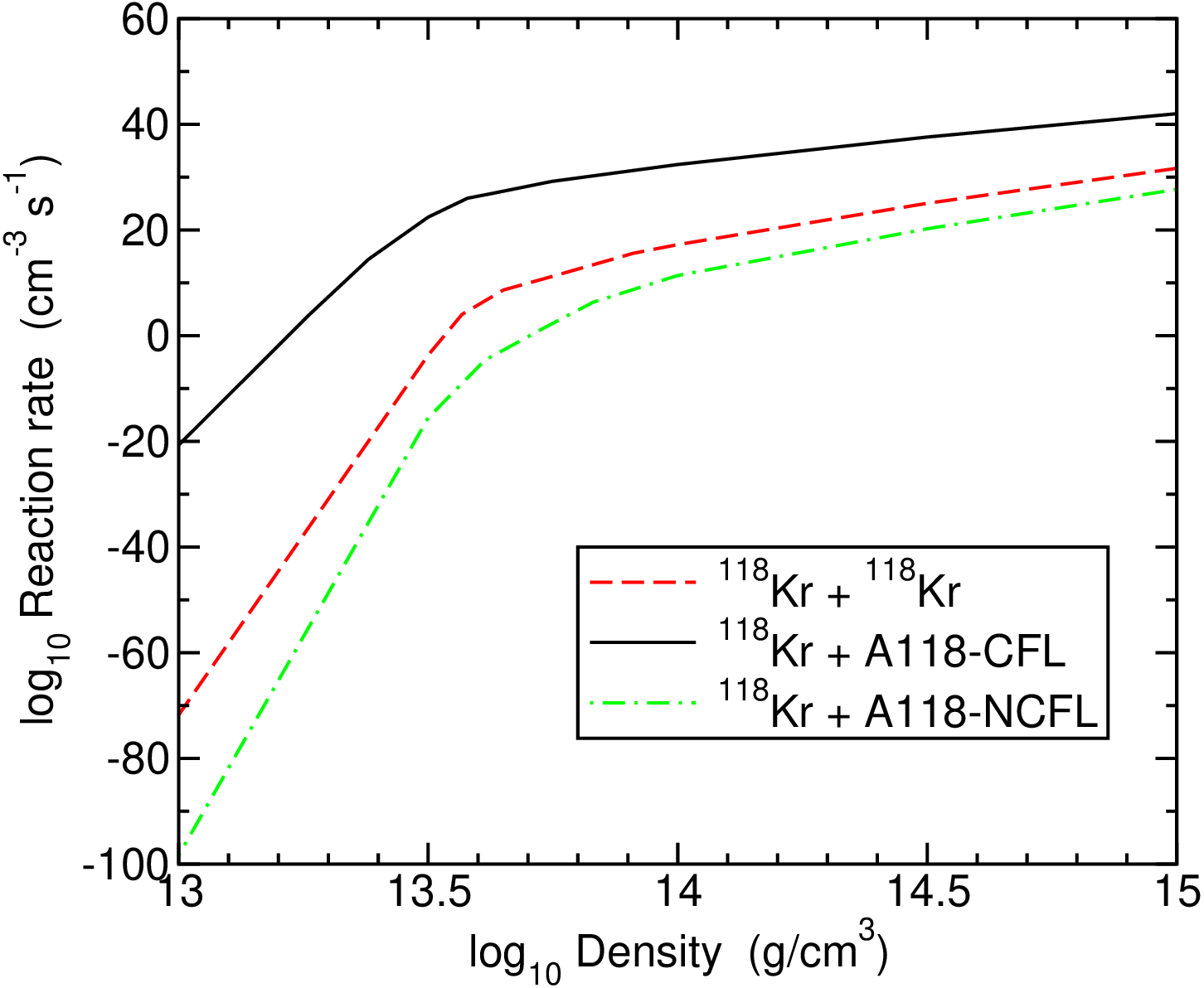}
  \caption{(Color online) Reaction rates for $^{118}$Kr and a
    strange quark matter nugget with 118 baryons as a function of
    density.}
\label{fig10}
\end{figure}

\begin{figure}[htb]
\centering\includegraphics[width=0.45\textwidth]{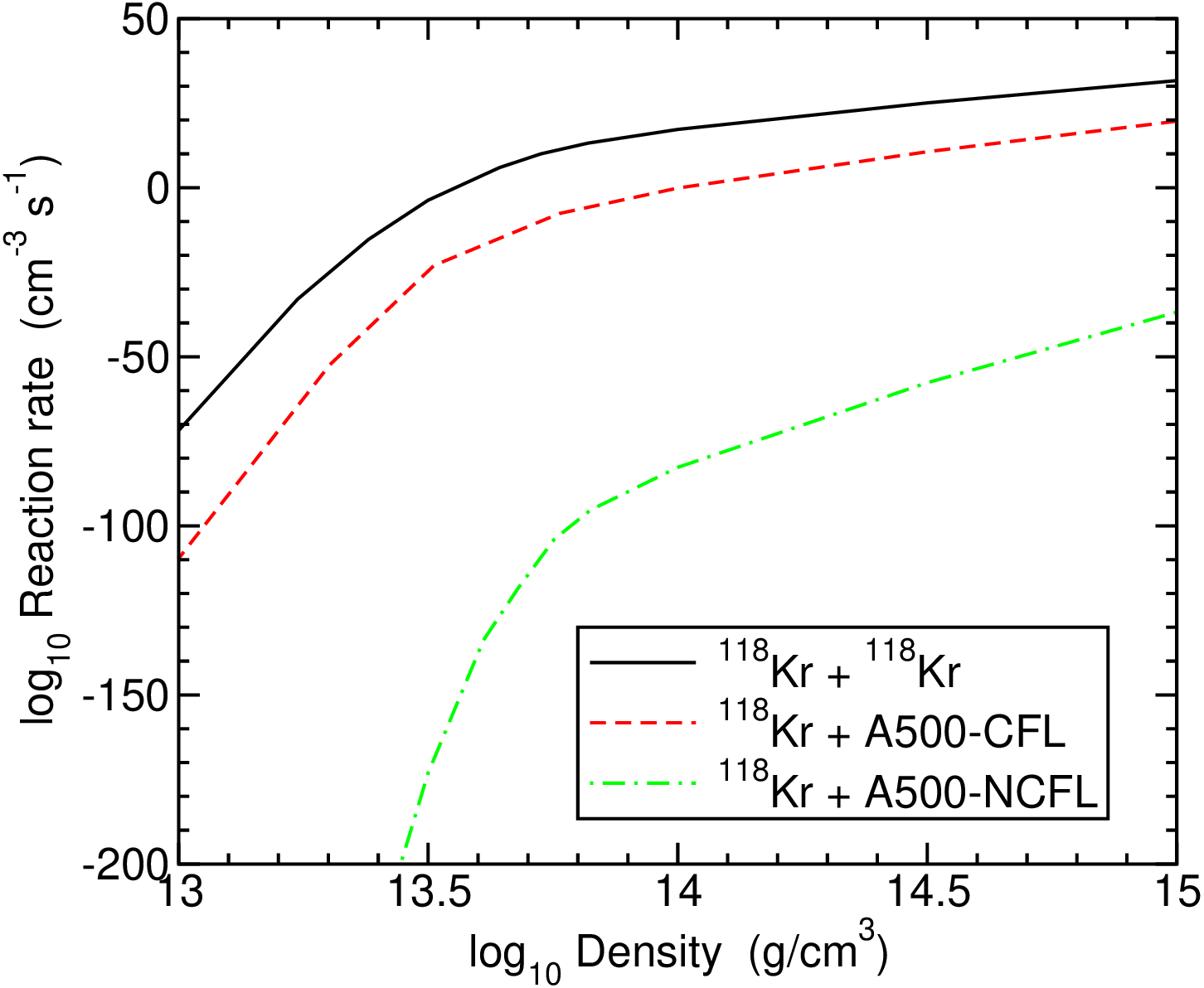}
\caption{(Color online) Reaction rates for $^{118}$Kr and a strange
  quark matter nugget with 500 baryons as a function of density.}
\label{fig11}
\end{figure}

After incorporating these changes due to strange quark matter,
pycnonuclear reaction rates between normal nuclear matter and matter
nuggets that incorporate some characteristics of strange quark matter
are compared in Figs. \ref{fig7} to \ref{fig11}. In Figs. \ref{fig7}
and \ref{fig10}, the atomic nuclei are paired with strangelets of
equivalent baryon number; in Figs. \ref{fig8} and \ref{fig11},
strangelets of baryon number 500 are used. Previously-shown data for
the reaction rates between two atomic matter nuclei are also provided
in the graphs for comparison. In these calculations, the strange quark
was assumed to have a mass of 300 MeV.

Again looking for possible trends, one can see that the pycnonuclear
reaction rates still decrease as the atomic numbers of the nuclei or
strangelets involved increase.  Additionally, the reaction rates
between atomic nuclei and the strange quark matter nuggets are very
different than that of normal nuclear matter alone. However, the
possibility of a color superconducting state for strange quark matter
has dramatic consequences. The extent of the increase in reaction
rates when compared to those of atomic nuclei alone is noteworthy,
with escalations of over twenty orders of magnitude, at a minimum, at
most densities of interest in the cases studied. By comparison,
non-CFL (NCFL) strange quark matter reaction rates are always
significantly lower than those of CFL strange quark matter with
equivalent baryon number, because of the electric charge
difference. The combination of these divergent effects -- with CFL
and non-CFL strange quark matter rates maintaining a significant gap
between them, the inclusion of CFL strange quark matter increasing the
reaction rates of atomic nuclei, and increasing baryon number
decreasing reaction rates for all systems -- makes it difficult to
determine {\em a priori} how reaction rates for a given system will
compare to each other.

\subsection{Quark Nugget Concentrations in the Crusts of
  Neutron Stars}

An upper limit on the quark nugget concentration in the crustal
regions of neutron stars accreting matter from white dwarfs would be
set by the quark nugget flux levels that could have been left over
from the Big Bang or from collisions of strange stars
\cite{witten,madsen98:b,cho94:a,cottingham94:a}.  The capture of these
quark nuggets by main-sequence stars would be an inevitable
consequence of the strange quark matter hypothesis. Due to their large
radii, main-sequence stars arise as large-surface
long-integration-time detectors for the quark nugget flux
\cite{madsen88:a,madsen89:a}. The accreted nuggets should be thermally
distributed throughout the interior of main-sequence stars, leading to
quark nugget concentrations that would monotonically grow over the
lifetime of main-sequence stars. For a constant bath of isotropic
monoenergetic quark nuggets, the baryon number accreted onto a
main-sequence star can be estimated from \cite{madsen88:a,madsen89:a}
\begin{eqnarray}
  A &\lsim& 1.6 \times 10^{47} \ \left(
    \frac{M}{\msun} \right)^{-0.15} \
  {v_{250}^{-1}} \ \rho_{24} 
  \nonumber \\
  && \left( 1 + 0.164 \ v_{250}^2 \
    \left( \frac{M}{\msun} \right)^{-0.25}\right)  \, ,
\label{eq:acmm}
\end{eqnarray} 
where $v_{250}\equiv v_\infty/ 250$~km/s and
$\rho_{24}\equiv\rho_\infty/ 10^{-24}~ {\rm g/cm}^3$. The quantities
$v_\infty$ and $\rho_\infty$ denote the quark nugget speed and
contribution to the density of the galactic halo far from the star,
respectively, and $M$ denotes the mass of a main-sequence star in
units of the mass of the sun, $\msun$.  If one assumes that all the
dark matter in the halo of our galaxy would consist of quark nuggets,
which is certainly an overestimate, then $\rovel\ \sim 1$. Equation
(\ref{eq:acmm}) then leads for typical progenitor star masses to $A
\lsim 10^{48}$ for the total baryon number of the quark nuggets in
white dwarfs, formed from main-sequence stars. This value is to be
compared with the typical baryon number of the crust of a neutron
star, which is around $10^{54}$. Hence, based on this scenario, one
would expect less than one quark nugget per $10^6$ atomic nuclei. Of
course, this ratio may be significantly greater in regions of the
Universe (e.g.\ Globular Clusters) where collisions of strange stars
would have generated quark-nugget flux levels that are greatly
enhanced in comparison to the homogeneously spread quark-nugget flux
level assumed above.

\section{Conclusions}\label{sec:conclusions}

The methodology presented in this paper provides insight into the
magnitude of the rates of pycnonuclear reactions between a simple
model of strange quark matter and atomic nuclei. Recreating a nuclear
interaction model and incorporating it in high-density nuclear
reaction processes, this effort reconstructs previous calculations for
normal nuclear matter. It then expands in scope, extending the model
to estimate the changes in the reaction rates as distinctive
theoretical characteristics of strange quark matter are substituted
for atomic matter.

Assessing the results for all of the data presented, there is an
unsurprising decrease in the reaction rates for nuclei with larger
atomic numbers, corresponding to the amplification of the Coulomb
barrier as the positive charge is increased. In the extended model,
though the charge suppression of strangelets was expected to have an
effect on pycnonuclear reaction rates, the increase by at least twenty
orders of magnitude for reactions between CFL strange quark matter and
atomic nuclei at the densities of a neutron star crust was
surprising. Reactions that incorporated non-CFL strangelets maintained
rates that were significantly lower than the CFL results, another
indicator of the additional reduction in charge for the color
superconducting system. With these conflicting trends and the
associated inability to presciently determine the reaction rates of
the differing systems with respect to each other, this methodology for
calculating pycnonuclear reaction rates will be of great importance
for future astrophysical studies.

As an example, we mention the incorporation of the fusion reaction
rates computed in this paper into pycnonuclear reaction fusion network
calculations, such as performed by JINA \cite{JINA}. Such calculations
will provide key information about the heat release from neutron stars
whose crusts may be contaminated with (superconducting) strange quark
matter nuggets. Since the reaction rates of the latter differ by many
orders of magnitude from the reaction rates of ordinary nuclear crust
matter, nuclear fusion network calculations may have the potential to
shed light on the existence of strange quark matter nuggets and on
whether or not they are in a color superconducting state, as suggested
by QCD
\cite{rajagopal01:a,alford01:a,alford98:a,rapp98+99:a,alford08:a}.

\subsection{Uncertainties}\label{ssec:uncertainties}

There are several areas of uncertainty. First, the pycnonuclear
reaction rate equation in Section \ref{sec:pycno} makes use of the WKB
integrals {\em J} and {\em K}, integrals that were numerically
calculated for densities between $1.7 \times 10^{-5} \tn{g cm}^{-3}
\le \rho/\langle A\rangle A^3 Z^6 \le 1.7 \times 10^4 \tn{g cm}^{-3}$
\cite{salpeter}. The densities possible from the outer crust to the
core of a neutron star are all several orders of magnitude greater
than this density range. In the attempt to recreate previous efforts
that used and did not change the pycnonuclear reaction rate equation,
the applicability of the Salpeter and Van Horn calculation at these
densities has been assumed.

There is also the issue of the nuclear potential model for atomic
matter. Referring again to Fig. \ref{fig2}, it should be noted that
the Sao Paulo model does not implement a repulsive, hard core for the
nuclear interaction. Given its focus on recreating nonlocal effects
with a local-equivalent potential, it is quite possible that it was
never intended to do so.  Though it shows similarities to the M3Y
potential in the surface region of nuclei and provides a good fit to
data of heavy-ion potential strengths \cite{globalbp}, its validity
has been called into question \cite{misicu,norepcore}. It is also
critical to remember that this study leaves the nuclear interaction
model unchanged when the adjustments for strange quark matter are
made. Any possible conclusions must be weighed against this
deficiency.

\subsection{Future Work}

There are several avenues available for future work. The initial
suggestions begin with the known sources of error just mentioned in
Sect.\ \ref{ssec:uncertainties}. Though any analysis and modification
of the {\em J} and {\em K} WKB integrals for the pycnonuclear reaction
rate equation is a task heavily steeped in numerical and complex
analysis, Salpeter and Van Horn do offer many details about their
calculations in their published work \cite{salpeter,vswkb}. A
verification that the current reaction rate equation properly accounts
for the densities under consideration in a neutron star would be of
use.

There is another opportunity in that the modularity of this
methodology allows for comparisons of different nuclear force
models. If one does not intend to model exotic matter, the current
methodology may still be used to assess changes in reaction rates due
to various nuclear interaction models. It may be possible that such
comparisons may provide support for the use of one nuclear model over
another.

Future efforts may include an improved model of the nuclear
interaction with strange quark matter. The current double-folding
process that determines the total nuclear potential uses a model of
the nuclear force that is assumed to be applicable to both the target
and projectile nuclei within the lattice. If a strangelet interaction
model is included, future efforts may need to alter the subordinate
nuclear potential $v_{NN}$ prior to its integration with the
strangelet and atomic matter densities. On the other hand, a simpler
analysis could focus on any effects of the reaction rates that may
stem from changing the mass of a strange quark, assumed to be 300~MeV,
within the current strange quark matter model.

Separately, it may also be useful to assess the effect of strangelets
or different atomic nuclei on the pycnonuclear reaction rates of
nearby interacting particle pairs. This sort of calculation, focused
on the effects of lattice imperfections, has not been extensively
studied even for normal nuclear matter. This is not a three-body
calculation, since the environment already contains many nearby
particles frozen in a lattice. The lattice imperfection would first
alter the geometry of the electrostatic screening potential in Section
\ref{sec:pycno}, affecting the vibrational energy levels of a nearby
interacting ion pair and, in turn, influencing the reaction
rate. Since the screened Coulomb potential is also used to determine
the wavefunction of the incoming particle, it is likely that this
particular line of research would require extensive changes to the
pycnonuclear reaction rate equation itself.  The methodology for the
calculation of the S-factor could remain unchanged.

In addition, follow-on efforts may wish to incorporate pycnonuclear
reaction rates into stellar thermal models, possibly flowing potential
characteristics of strange quark matter to observable predictions. As
a first approximation, the zero-temperature pycnonuclear reaction rate
was used in this effort; future efforts may wish to incorporate the
temperature-dependent pycnonuclear reaction rate equation developed by
Salpeter and Van Horn \cite{salpeter}.  One could build on this by
calculating an energy generation rate in units of ${\rm erg}~ {\rm
  g}^{-1} ~ {\rm s}^{-1}$ that can be obtained by multiplying the
pycnonuclear reaction rate equation by $Q/2 \rho$
\cite{salpeter}. {\em Q} is the energy release in a single
reaction. Approximations for {\em Q} would need to be made for strange
quark matter, perhaps building on estimates that the total energy
release for normal nuclear matter is 1.45 MeV per baryon
\cite{thermal}. These thermal calculations could then be incorporated
into stellar heating models, evaluating possible variations in total
heat output for neutron stars due to normal matter or strangelets that
may be present as a result of state changes or accretion. Differences
in reaction rates may thus provide a clue to the presence of strange
quark matter in astrophysical environments.

In any case, it is hoped that this investigation may serve as a
stepping stone for further studies, and that these future
possibilities may tie astrophysical observations of high-density
objects to speculations about the possible existence of (color
superconducting) strange quark matter.

\section*{Acknowledgments}

This work was supported by the National Science Foundation (USA) under
Grant PHY-0457329.


\begin{thebibliography}{9}

\bibitem{glen97:book} N.K. Glendenning, {\it Compact Stars, Nuclear
    Physics, Particle Physics, and General Relativity},
  Springer-Verlag, New York, 2000.

\bibitem{weber99:book} F. Weber, {\it Pulsars as Astrophysical
    Laboratories for Nuclear and Particle Physics}, High Energy
  Physics, Cosmology and Gravitation Series, IOP Publishing, Bristol,
  Great Britain, 1999.

\bibitem{ivanenko65:a} D.D. Ivanenko and D.F. Kurdgelaidze,
  Astrophys.\ {\bf 1}, 251 (1965).

\bibitem{fritzsch73:a} H. Fritzsch, M. Gell--Mann, and H. Leutwyler,
  Phys. Lett. {\bf 47B}, 365 (1973).

\bibitem{baym76:a} G. Baym and S.Chin, Phys.\ Lett.\ {\bf 62B}, 241 (1976).

\bibitem{keister76:a} B.D. Keister and L.S. Kisslinger, Phys.\
  Lett.\ {\bf 64B}, 117 (1976).

\bibitem{chap77:a+b} G. Chapline and M. Nauenberg, Phys.\ Rev.\ D {\bf
    16}, 450 (1977); Ann.\ New York Academy of Sci.\ {\bf 302}, 191
  (1977).

\bibitem{fech78:a} W.B. Fechner and P.C. Joss, Nature {\bf 274}, 347
  (1978).

\bibitem{glen91:pt} N.K. Glendenning, Phys.\ Rev.\ D {\bf 46}, 1274
  (1992) 1274.

\bibitem{alford07:nature} M. Alford, D. Blaschke, A. Drago, T. Kl{\"{a}}hn,
  G. Pagliara, and J. Schaffner-Bielich, Nature {\bf 445}, 7 (2007).

\bibitem{kaplan86:a} D.B. Kaplan and A.E. Nelson, Phys.\ Lett.\ {\bf
    175B}, 57 (1986); {\it ibid.}  Nucl.\ Phys. {\bf A479}, 273 (1988).

\bibitem{brown87:a} G.E. Brown, K. Kubodera, and M. Rho, Phys.\
  Lett.\ {\bf 192B}, 273 (1987).

\bibitem{brown95:a} G.E. Brown, {\it Kaon condensation in dense
    matter}, in: Bose--Einstein Condensation, ed.\ A.\ Griffin, D.\
  W.\ Snoke, and S.\ Stringari, Cambridge Univ.\ Press, 1995, p.\ 438.

\bibitem{li97:ab} G.Q. Li, C.-H. Lee, and G.E. Brown, Nucl.\ Phys.\
  {\bf A625}, 372 (1997); {\it ibid.} Phys.\ Rev.\ Lett.\ {\bf 79}
  (1997) 5214.

\bibitem{brown97:a} G.E. Brown, Phys.\ Bl.\ {\bf 53}, 671 (1997).

\bibitem{lattimer01:a} J.M. Lattimer and M. Prakash, Astrophys.\ J.\
  {\bf 550}, 426 (2001).  

\bibitem {weber05:a} F. Weber, Progress in Particle and Nuclear Physics
  {\bf 54}, 193 (2005).

\bibitem{klaehn06:a} T. Kl{\"{a}}hn {\it et al.}, Phys. Rev. C
    {\bf 74}, 035802 (2006).

\bibitem{sedrakian07:a} A. Sedrakian, Prog.\ Part.\ Nucl.\ Phys.\ {\bf
    58}, 168 (2007).

\bibitem{page06:a} D. Page and S. Reddy, Ann. Rev. Nucl. Part.
  Sci. {\bf 56}, 327 (2006).

\bibitem{farhi84:a} E. Farhi and R. Jaffe, Phys. Rev. D {\bf 30}, 2379
  (1984).

\bibitem{schaffner97:a} J. Schaffner-Bielich, C. Greiner, A. Diener,
  and H. Stoecker, Phys. Rev. C {\bf 55}, 3038 (1997).

\bibitem{alcock86:a} C. Alcock, E. Farhi, and A. Olinto,
  Astrophys. J. {\bf 310}, 261 (1986).

\bibitem{alcock88:a} C. Alcock and A. V.  Olinto,
  Ann. Rev. Nucl. Part. Sci. {\bf 38}, 161 (1988).

\bibitem{madsen98:b} J. Madsen, Lecture Notes in Physics {\bf 516},
  162 (1999).  

\bibitem {witten} E. Witten, Phys. Rev. D {\bf 4}, 272 (1984).

\bibitem {bodmer} A.R. Bodmer, Phys. Rev. D {\bf 4}, 1601 (1971).

\bibitem {terazawa} H. Terazawa, INS-Report-338 (INS, Univ. of Toyko,
  1979); J. Phys. Soc. Japan, {\bf 58}, 3555 (1989); {\it ibid.} {\bf 58},
  4388 (1989); {\it ibid.} {\bf 59}, 1199 (1990).

\bibitem {rajagopal01:a} K. Rajagopal, F. Wilczek, in: M. Shifman (Ed.),
  {\em The Condensed Matter Physics of QCD, At the Frontier of
    Particle Physics/Handbook of QCD}, World Scientific, 2001.

\bibitem{alford98:a} M. Alford, K. Rajagopal, and F. Wilczek, Phys.\
  Lett.\ {\bf 422B}, 247 (1998).


\bibitem{alford01:a} M. Alford, Ann. Rev. Nucl. Part. Sci. {\bf 51},
  131 (2001).

\bibitem{rapp98+99:a} R. Rapp, T. Sch{\"{a}}fer, E.V. Shuryak, and
  M. Velkovsky, Phys.\ Rev.\ Lett.\ {\bf 81}, 53 (1998); {\it ibid.}
  Ann.\ Phys.\ {\bf 280}, 35 (2000).

\bibitem{alford08:a} M.G. Alford, A. Schmitt, K. Rajagopal, and
  T. Sch{\"{a}}fer, Rev. Mod. Phys. {\bf 80}, 1455 (2008).


\bibitem{berger87:a} M.S. Berger and R.L. Jaffe, Phys. Rev. C {\bf
    35}, 213 (1987).

\bibitem{schaffner92:a} J. Schaffner, C. Greiner, and
  H. St{\"{o}}cker, Phys. Rev. C {\bf 46}, 322 (1992).

\bibitem{gilson93:a} E.P. Gilson and R.L. Jaffe, Phys. Rev. Lett.
  {\bf 71}, 332 (1993).

\bibitem{zhang02:a} Y. Zhang and R.-K. Su, Phys. Rev. C {\bf 67},
  (2003) 015202.

\bibitem {peierls} R.E. Peierls, {\em Quantum Theory of Solids},
  Oxford University Press, Oxford, 2001.

\bibitem {shapiro} S.L. Shapiro and S.A. Teukolsky, {\em Black Holes,
    White Dwarfs, and Neutron Stars.} John Wiley \& Sons, Inc., New
  York, 1983.

\bibitem {salpeter} E.E. Salpeter and H.M. Van Horn,
  Astrophys. J. {\bf 155}, 183 (1969).

\bibitem {ocppycno} L.R. Gasques, A.V. Afanasjev, E.F. Aguilera,
  M. Beard, L.C. Chamon, P. Ring, M.  Wiescher, and D.G. Yakovlev,
  Phys. Rev. C {\bf 72}, 025806 (2005).
  
\bibitem {clayton} D.D. Clayton, {\em Principles of Stellar Evolution
    and Nucleosynthesis}, University of Chicago Press, Chicago,
  1983.  
  
\bibitem {mcppycno} D.G. Yakovlev, L.R. Gasques, A.V. Afanasjev,
  M. Beard, and M. Wiescher, Phys. Rev. C {\bf 74}, 035803 (2006).

\bibitem {carr} W.J. Carr, Jr., Phys. Rev. {\bf 122}, 1437 (1961).  
  
\bibitem {globalbp} L.C. Chamon, B.V. Carlson, L.R. Gasques,
  D. Pereira, C. De Conti, M.A.G. Alvarez, M.S. Hussein,
  M.A.C. Ribeiro, E.S. Rossi, Jr., and C.P. Silva, Phys. Rev. C {\bf
    66}, 014610 (2002).

\bibitem {bpdatachk} L.R. Gasques, L.C. Chamon, D. Pereira,
  M.A.G. Alvarez, E.S. Rossi, Jr., C.P. Silva, and B.V. Carlson,
  Phys. Rev. C {\bf 69}, 034603 (2004).
  
\bibitem {braz} L.C. Chamon, B.V. Carlson, L.R. Gasques, D. Pereira,
  C. De Conti, M.A.G. Alvarez, M.S. Hussein, M.A.C. Ribeiro,
  E.S. Rossi, Jr., and C.P. Silva, Brazilian Journal of Physics {\bf
    32}, 238 (2003).  

\bibitem {paulinonloc} M.A.C. Ribeiro, L.C. Chamon, D. Pereira,
  M.S. Hussein, and D. Galetti, Phys. Rev. Lett. {\bf 78}, 3270
  (1997).
  
  \bibitem {lett79} L.C. Chamon, D. Pereira, M.S. Hussein,
  M.A.C. Ribeiro, and D. Galetti, Phys. Rev. Lett. {\bf 79}, 5218
  (1997).
  
  \bibitem {c58} L.C. Chamon, D. Pereira, and M.S. Hussein,
  Phys. Rev. C {\bf 58}, 576 (1998).

\bibitem {bparam} D.F. Jackson and R.C. Johnson, Phys. Lett. {\bf
    49B}, 249 (1974).

\bibitem {perey} F. Perey and B. Buck, Nucl. Phys. {\bf 32}, 253
  (1962).

\bibitem {schiff} L.I. Schiff, {\em Quantum Mechanics}, 3rd
  ed. McGraw-Hill, New York, 1968.
  
\bibitem {madseng28} J. Madsen, J. Phys. {\bf G28}, 1737 (2002).

\bibitem{rajagopal01:PRL} K. Rajagopal and F. Wilczek,
  Phys. Rev. Lett. {\bf 86}, 3492 (2001).

\bibitem {madsenlett87} J. Madsen, Phys. Rev. Lett. {\bf 87}, 172003 (2001).

\bibitem {oertel08:a} M. Oertel and M. Urban, Phys. Rev. D \ {\bf
    77}, 074015 (2008).

\bibitem{cho94:a} S.J. Cho, K.S. Lee, and U. Heinz, Phys.\ Rev.\ D
  {\bf 50}, 4771 (1994).

\bibitem{cottingham94:a} W. N. Cottingham, D. Kalafatis, and R. Vinh
  Mau, Phys. Rev. Lett.\ {\bf 73}, 1328 (1994).

\bibitem{madsen88:a} J. Madsen, Phys. Rev. Lett. {\bf 61}, 2909 (1988).

\bibitem{madsen89:a} J. Madsen, {\it Quark Nuggets, Dark Matter and
    Pulsar Glitches}, Proc.\ of the XXIVth Rencontre de Moriond, Les
  Arcs, Savole, France, March 5--12, 1989, ed.\ by J.\ Audouze and J.\
  Tran Thanh Van (Edition Frontiers, 1990).

\bibitem {misicu} S. Misicu and H. Esbensen, Phys. Rev. C {\bf 75},
  034606 (2007).

\bibitem {norepcore} L.R. Gasques, E. Brown, A. Chieffi, C.L. Jiang,
  M. Limongi, C. Rolfs, M. Wiescher, and D. G. Yakovlev,
  Phys. Rev. C {\bf 76}, 035802 (2007).

\bibitem {vswkb} H.M. Van Horn and E.E. Salpeter, Phys. Rev. {\bf 157},
  751 (1967).

\bibitem {thermal} D.G. Yakovlev, K.P. Levenfish, and O.Y. Gnedin,
  Eur. Phys. J. {\bf A25}, 669 (2005).

\bibitem {JINA} Information about JINA can be found at {\tt
    www.jinaweb.org}.

\end{thebibliography}
\end{document}